\documentclass[11pt]{article}

\usepackage{amsmath}
\usepackage{graphicx}
\usepackage{amssymb}
\usepackage{amsfonts}

\def\beq{\begin{eqnarray}}
\def\eeq{\end{eqnarray}}
\def\ln{\,\mbox{ln}\,}

\def\al{\alpha}
\def\be{\beta}

\def\de{\delta}
\def\vp{\varepsilon}

\def\la{\lambda}
\def\na{\nabla}
\def\pa{\partial}

\def\ph{\varphi}

\def\Ga{\Gamma}

\def\La{\Lambda}

\begin{document}

\begin{center}
{\large\sc  On the renormalization of gauge theories in curved
space-time.}
\vskip 6mm

Peter M. Lavrov$^{a}\footnote{E-mail address: lavrov@tspu.edu.ru}$,
\quad
Ilya L. Shapiro$^{b}\footnote{Also at Tomsk State Pedagogical
University, Tomsk, Russia. \ E-mail address: shapiro@fisica.ufjf.br}$
\vskip 6mm

(a) \ \ {\small\sl Department of Mathematical Analysis, \\Tomsk State
Pedagogical University,\\ 634061, Kievskaya St. 60, Tomsk, Russia}
\vspace{0,4cm}

(b) \ \ {\small\sl Departamento de F\'{\i}sica, ICE,\\
Universidade Federal de Juiz de Fora
\\
Juiz de Fora, CEP: 36036-330, MG,  Brazil}

\end{center}

\begin{quotation}
\noindent
{\large\bf Abstract.}
We consider the renormalization of general gauge theories on 
curved space-time background, with the main assumption being 
the existence of a gauge-invariant and diffeomorphism invariant
regularization. Using the  Batalin-Vilkovisky (BV) formalism 
one can show that the theory possesses gauge invariant and diffeomorphism invariant renormalizability at quantum level, 
up to an arbitrary order of the loop expansion. Starting from 
this point we discuss the locality of the counterterms and the 
general prescription for constructing the power-counting 
renormalizable theories on curved background.
\vskip 4mm

\vskip 2mm

\vskip 2mm

\noindent
{\sl MSC-AMS:} \
81T20, \
81T15, \

\end{quotation}
\vskip 12mm

\section{Introduction}

The quantum field theory (QFT) in curved space is an
important ingredient of our general understanding of the
quantum description of nature. The reason for this is that,
according to General Relativity our space-time is likely
to be curved. Therefore, as far as we think
that the QFT approach is a fundamental one in the description
of the interaction of elementary particles and fields, it
must be considered on a curved space-time background.
The consideration of QFT on classical curved background
does not rule out the quantization of gravity, but, in
some sense, is at least equally important. The reason is
that we do not know which one of the existing ways to
quantize gravity is close to reality, while the QFT of
matter fields definitely deals with reality, as the
concept of a classical curved space does.

One of the most important aspects of the modern QFT is the theory of
gauge fields and their perturbative renormalization. The gauge
invariant renormalizability is the corner stone in the construction
of the very important theories including the Standard Model of
particle physics. Hence it is quite interesting to know whether the
existing methods to analyze renormalizability of gauge theories are
working well in curved space. In the previous considerations of the
problem \cite{buch84,Toms} (see also \cite{Panangaden}) it has been
assumed that the gauge invariant renormalization of the theory is
indeed possible, due to the existence of both gauge invariant and
diffeomorphism invariant regularization, such as a dimensional one. 
Starting from this point it is possible to
establish the prescription for constructing the renormalizable
theories of interacting matter fields on curved background
\cite{buch84,book} (see also \cite{PoImpo} for a recent review and
for somehow more simple treatment of the issue).

The present work is intended to explore, in a more formal way than 
it was done before, the issue of gauge invariant renormalizability 
in curved space-time. For this end we are going to apply the BV 
formalism. It is well known that this formalism enables one to 
prove the gauge-invariant renormalizability of general gauge
theories in a situation when all fields under consideration 
are quantum ones
\cite{VLT,GomisWein} (see also \cite{HT,ABRKT} for an extensive
review and further references).  It is of course important to
generalize these considerations to the case when the QFT is defined
in the presence of external conditions, in particular in curved
space-time. In this case one has to take care about both gauge 
symmetries and general covariance. The last symmetry involves 
both quantum and external fields, making the consideration 
more complicated. Our main purpose is to consider the general 
features of renormalization of the theory of quantum matter 
fields in curved space-time, using the powerful BV formalism. 
On the top of that we will discuss the construction of
multiplicatively  renormalizable theories in curved space, the 
subject which was already considered previously (see, e.g. 
\cite{book,PoImpo} and references therein) in a slightly 
different manner.  

The paper is organized as follows. In the next section we 
present a very brief review of the antibracket (BV) formalism
in gauge theories. In Sect. 3 we consider the same 
formalism for gauge theories in curved space. The gauge-invariant 
renormalization in curved space-time is considered in Sect. 4. 
An important aspect of the theory is the possibility to use 
the non-covariant gauge fixing conditions, which is discussed 
in Sect. 5. In Sect. 6 we introduce the quantum gravity 
completion of the theory to get some strong arguments 
supporting the locality of the counterterms of the quantum 
theory in curved space. The power-counting renormalizability 
and the receipt for constructing renormalizable theories in 
curved space are discussed in Sect. 7. Finally, in Sect. 8 
we draw our conclusions.

\section{Gauge theories in BV formalism}

In this section we present a very brief review of the
BV  formalism \cite{BV}, which
will be used in the rest of the paper to prove the gauge
invariant and general covariant renormalizability of the
quantum field theory on curved background. An extensive
review of the formalism can be found in \cite{HT,ABRKT}, here
we mainly collect information (and also fix notations) which
will be needed in further consideration.

\subsection{Preliminaries and terminology}

The need for the advanced version of the Lagrangian
quantization formalism was inspired by the discovery of
supergravity theories in 70-ies \cite{Sugra}.
The gauge transformations possess linearly-dependent
generators and, as a consequence, direct application
of the Faddeev-Popov procedure leads to the violation
of unitarity of the physical $S$-matrix.
Moreover, attempts of covariant quantization of  gauge
theories with linearly-dependent generators of gauge
transformations result in the understanding of the fact
that it is impossible to use the Faddeev-Popov rules to
construct a suitable quantum theory \cite{Lindep}.
The quantization of general gauge theories requires taking
into account such aspects as the existence of open algebras
and reducible generators. The quantization can be performed
only by introducing different types of ghosts, antighosts,
ghosts for ghosts (Nielsen, Kallosh ghosts etc.)
\cite{PrimBV}. A unique
closed approach to the problem of covariant quantization
summarized all these attempts was proposed by Batalin and
Vilkovisky \cite{BV}. The BV formalism
gives the rules for the quantization of general gauge
theories.

The starting point of the BV method is a theory of fields
$\,A^i \, (i=1,2,...,n)$ with Grassmann parities
$\,\vp(A^i)=\vp_i$, for which the
initial classical action $S_0(A)$ is assumed to have at
least one stationary point $A^{i}_{0}$
\beq
\label{EOMClassA}
S_{0,i}(A)|_{A_0}=0,\quad
\eeq
and to be regular in the neighborhood of $A_0$. Here we
are using the notations
$$
A_0=\{A^i_0\}\qquad \mbox{and}\qquad
F_{,i}(A) = \frac{\pa_r F(A)}{\pa A^i}\,,
$$
where the label ''$r$'' denotes the right derivative.

Geometrically, the Eqs. (\ref{EOMClassA}) define a surface
$\Sigma$ in the space of functions $A^i$. We assume
the invariance of the action $S_0(A)$ under the gauge
transformations $\delta A^i=R^i_{\alpha}(A) \xi^{\alpha}$
in the neighborhood of the stationary point,
\beq
\label{GIClassA}
 S_{0,i}(A)\, R^i_{\alpha}(A) = 0\,,\qquad
\alpha=1,2,...,m
\,,\qquad
0 < m < n
\,,\qquad
\vp(\xi^\al) = \vp_\al\,.
 \eeq
Here $\xi^{\alpha}$ are arbitrary functions of space-time
coordinates ,
and $R^i_{\alpha}(A)$
are generators of gauge transformations. We have also used DeWitt's
condensed notations \cite{DeW}, such that any index includes space -
time, index of internal group, Lorentz index and so on. 
Consequently, a summation over repeated indices includes, along with
summation over internal and Lorentz indices, also an integration
over continuous variables such as space-time coordinates.

It follows from the Noether identities (\ref{GIClassA}) 
that, first, the equations of motion are not independent 
and, second, (some) propagators do not exist because the 
Hessian matrix $H_{ij} = S_{0,ij}$ corresponding to the 
action $S_0$  is degenerate at any point on the 
stationary surface $\Sigma$,
\beq
\nonumber
 S_{0,i}(A) \,R^i_{\alpha ,j}(A) +
 S_{0,ji}(A)\,R^i_{\alpha} (-1)^{\vp_{\al}\vp_j} = 0
\quad \Longrightarrow \quad
S_{0,ji}\,R^i_{\alpha}|_{A_0} = 0\,.
\eeq
The generators $R^i_{\alpha}$ are on shell zero-eigenvalue
vectors of the Hessian matrix $S_{0,ij}$.

The structure of gauge algebra can be found by studying the
commutator of gauge transformations and some consequences
from the relations (\ref{GIClassA}). We assume that the set
of generators $\,R^i_{\alpha}(A)\,$ is complete. In this
case one can prove that the generators algebra has the
following general form (see \cite{VT1,BV3,BV4}):
\beq
\label{GAGGT}
R^i_{\alpha , j}(A)R^j_{\beta}(A)-(-1)^{\vp_{\alpha}\vp_
{\beta}}R^i_{\beta ,j}(A)R^j_{\alpha}(A)=-R^i_{\gamma}(A)
F^{\gamma}_{\alpha\beta}(A)- S_{0,j}(A)M^{ij}_{\alpha\beta}(A)\,,
\eeq
where $F^{\gamma}_{\alpha\beta}(A)$ are structure functions
with the following symmetry properties:
$$
F^{\gamma}_{\alpha\beta}(A)
\,=\,-\,(-1)^{{\vp_\al\vp_\be}}F^{\gamma}_{\be\al}(A)
$$
and $M^{ij}_{\alpha\beta}(A)$ are satisfying the conditions
$$
M^{ij}_{\alpha\beta}(A) = -(-1)^{\vp_i\vp_j}
M^{ji}_{\alpha\beta}(A) =
-(-1)^{\vp_{\alpha}\vp_{\beta}}M^{ij}_{\beta\alpha}(A)\,.
$$

In case \ $M^{ij}_{\alpha\beta}(A) = 0$, one meets a gauge
theory with a {\it closed} gauge algebra.
If $M^{ij}_{\alpha\beta}(A)\neq 0$, then
the gauge algebra is called {\it open}. In this case,
due to the symmetry properties of $M^{ij}_{\alpha\beta}(A)$,
the quantities
$$
R^i_{\al\be, triv}(A) = S_{0,j}(A)M^{ij}_{\al\be}(A)
$$
are symmetry generators of the initial
action $S_0(A)$ which can be called trivial. They vanish
at the extremals of $S_0(A)$,
\beq
R^i_{\alpha\beta, triv}(A)|_{S_{0, i}=0}=0
\nonumber
\eeq
and leave the action invariant. At the same time, they are
not connected with an additional degeneration of the
initial action $\,S_0(A)$, because the rank of the Hessian
matrix  describing the degeneracy of the initial
action, is defined at the extremals $S_{0,i}=0$.

Finally, if \ $M^{ij}_{\alpha\beta}(A) = 0$ \ and
\ $F^{\gamma}_{\alpha\beta}$ \ do not depend on the fields
\ $A$, the gauge transformations form a gauge group and
define a {\it Lie algebra}.

\subsection{BV quantization: the general procedure}

The procedure of the BV-quantization for a general
gauge theory involves the following steps. First, the total
configuration space of the fields $\phi^A$ is introduced. 
For irreducible theories the fields $\phi^A$ include $\,A^i$, 
ghost and antighost fields $C^{\alpha}$ and $\bar{C}^{\alpha}$ 
and auxiliary (Nakanishi-Lautrup) fields $B^{\alpha}$
\beq
\label{ConfSpaceBV}
\phi^A = (A^i,\;B^{\alpha},\;C^{\alpha},\;\bar{C}^{\alpha}),\quad
\vp (\phi^A) = \vp_A,
\eeq
with the following distribution of the Grassmann parities and
ghost numbers
\beq
\nonumber
\vp (A^i) = \vp_i\,,
\qquad
\vp (B^{\alpha}) = \vp_{\alpha}\,,
\qquad
\vp (C^{\alpha})  = \vp (\bar{C}^{\alpha})
= \vp_{\alpha}+1\,, \\
\nonumber
gh(A^i)=gh(B^{\alpha}) = 0\,,
\qquad
gh(C^{\alpha}) = 1 \,,
\qquad
gh(\bar{C}^{\alpha}) = - 1\,.
\nonumber
\eeq

To each field $\phi^A$ of the total configuration space,
one introduces corresponding {\it antifield} $\phi^*_A$,
\beq
\label{antiSpaceRBV} \phi^*_A &=& \Big(A^*_i,\;B^*_{{\alpha}},\;
C^*_{{\alpha}}, \; \bar{C}^*_{{\alpha}}\Big).
\eeq

The statistics of $\phi^*_A$ is opposite to the statistics of the
corresponding fields $\phi^A$
\beq
\vp (\phi^*_A) = \vp_A + 1
\nonumber
\eeq
and ghost numbers of fields and corresponding antifields are connected
by the rule
\beq
\nonumber
\quad gh(\phi^*_A)=-1 -gh(\phi^A)\,.
\eeq

On the space of the fields $\phi^A$ and antifields
$\phi^*_A$ one defines an odd symplectic structure $(\;,\;)$ called the
 antibracket
\beq
\label{DefAB}(F, G)\,
\equiv \,
\frac{\delta F}{\delta\phi^A}\,\frac{\delta G}{\delta\phi^*_A}
- (F\leftrightarrow G)\;
(-1)^{\left[\vp(F)+1\right]\cdot\left[\vp(G)+1\right]}\,\,.
\eeq
Here the derivatives with respect to fields are understood as the
right ones and those with respect to antifields as the left ones.

One can easily verify that the following properties of the
antibracket follow from the definition (\ref{DefAB}):
\\
1) Grassmann parity relations
\ $\vp((F, G))=\vp(F)+\vp(G)+1=\vp((G, F))$;
\\
2)  Generalized antisymmetry
\ $(F, G)=-(G, F)(-1)^{(\vp(F)+1)(\vp(G)+1)}$;
\\
3)  Leibniz rule
\ $(F, GH)=(F, G)H+(F, H)G(-1)^{\vp(G)\vp(H)}$;
\\
4) Generalized Jacobi identity
\ $((F, G), H)(-1)^{(\vp(F)+1)(\vp(H)+1)}
+{\sf cycle} (F, G, H)\equiv 0\,.$
\vskip 3mm

Furthermore, one can readily check that the antibracket
(\ref{DefAB}) is invariant under the {\it anticanonical}
transformation of variables $\phi, \phi^*$ with the generating
functional \ $X=X(\phi,\phi^*)$, \ $\vp(X)=1$,
\beq
\label{AntiCT}
\phi^{'A}= \frac{\delta X(\phi,\phi^{*'})}{\delta\phi^{*'}_A},\quad
\phi^*_A=\frac{\delta X(\phi,\phi^{*'})}{\delta\phi^A} .
\eeq This property of the odd symplectic structure (\ref{DefAB}) on
the space of $\phi,\;\phi^*$ is a counterpart to the invariance
property of the even symplectic structure (the Poisson bracket)
under a canonical transformation of canonical variables ($p,\;q$).
For the first time, the importance of anticanonical transformations
(\ref{AntiCT}) in the formulation of the BV-method was realized in
\cite{VLT}.

As a second step the nilpotent generating operator $\Delta$ is
introduced according to
\beq
\label{DeltaBV}
\Delta=(-1)^{\vp_A} \frac{\delta_{\it l}}
{\delta\phi^A}\;\frac{\delta} {\delta\phi^*_A}
\,,\qquad
{\Delta}^2=0\,,
\qquad
\vp (\Delta)=1\,.
\eeq
We will always assume that formal manipulations with operators 
such as $\Delta$ can be supported by suitable regularization
scheme. This is a nontrivial requirement, since the operator
(\ref{DeltaBV}) is not well-defined on local functionals. The
reason is that for any local functional $S$, \
$\Delta S\sim\delta(0)$ and one faces the so-called problem
of $\delta(0)$. The usual way to deal with this problem is to
use the dimensional regularization \cite{Leib}, where
$\delta(0)$ is equal to zero.  Recently, a new calculus for
local variational differential operators in local quantum field
theory has been proposed by Shahverdiev, Tyutin and Voronov
\cite{STV}, where $\delta(0)$ does not arise at all.

Note that acting by $\Delta$ on the product of two functionals
$F$ and $G$ reproduces the antibracket,
\beq
\nonumber
\Delta [F\cdot G]=(\Delta F)\cdot G + F\cdot (\Delta G)
(-1)^{\vp (F)} + (F,G)(-1)^{\vp (F)}\,.
\eeq

As a third step the quantum master equation is defined according to
\beq
\label{MastEBV}
\frac {1}{2} (S,S)=i\hbar{\Delta}S
\eeq
or, equivalently,
\beq
\label{MastEqBV}
{\Delta}\exp\bigg\{\frac{i}{\hbar}S\bigg\}=0,
\eeq
where $S=S(\phi,\phi^*)$ is a bosonic functional satisfying the boundary
condition
\beq
\label{BoundCon}
S|_{\phi^* = \hbar = 0}= S_0(A).
\eeq
The bosonic functional $S$ is the fundamental object of the
BV-quantization scheme.
\vskip 3mm

The generating functional of Green's functions $Z(J)$ is defined as
\beq
\label{ZBV}
\nonumber
Z(J) &=& \int d\phi \exp \Big\{\frac{i}{\hbar}
\big[S_{eff}(\phi)+ J_A\phi^A\big]\Big\}\,,
\\
S_{eff}(\phi) &=&
S\Big(\phi,\,\phi^* \,=\, \frac{\delta\Psi}{\delta\phi}\Big)\,.
\label{Seff}
\eeq
Here $\Psi = \Psi(\phi)$ is a fermionic gauge
functional. For instance, if the gauge fixing condition in the 
Yang-Mills theory is chosen to be $\chi_{\alpha}=0$, the fermionic 
gauge functional has the form $\Psi={\bar c}^{\alpha}\chi_{\alpha}$, 
where ${\bar c}^{\alpha}$ is the FP antighost. Furthermore, in the
Eq. (\ref{Seff}), $J_A$ are the usual external sources to the
fields $\phi^A$. The Grassmann parities of these sources are
defined in a natural way, \ $\vp(J_A) = \vp_A$.

Note \cite{VLT}, that the gauge-fixing procedure (\ref{ZBV})
in the BV-quantization can be described in terms of anticanonical
transformation of the variables $\phi,\phi^*$ (\ref{AntiCT}) in
$S(\phi,\phi^*)$ with the generating functional $X$
\beq
\label{}
X(\phi,\phi^{*}) = \phi^{*}_A\phi^A + \Psi(\phi).
\nonumber
\eeq

To discuss some features of the BV-quantization, it is convenient to
rewrite the expression for the generating functional $Z(J)$ in the
equivalent form
\beq
\label{GFZUFBV}
\nonumber
Z(J) &=& \int d\phi \,d\phi^*\,
\delta \Big(\phi^* - \frac{\de\Psi}{\de\phi} \Big)
\exp\left\{\frac{i}{\hbar}[S(\phi,\phi^*)
+ J_A\phi^A]\right\}
\\
&=& \int d\phi \,d\phi^* \,d\lambda
\exp\left\{\frac{i}{\hbar}\bigg[S(\phi,\phi^*)
+ \Big(\phi^*_A 
- \frac{\de\Psi}{\de\phi^A}\Big)\lambda^A
+ J_A\phi^A\bigg]\right\}\,,
\eeq
where we have introduced the auxiliary (Nakanishi-Lautrup) fields
$\lambda^A$ \ with \ $\vp(\la^A) = \vp_A + 1$.

Note, first of all, that the integrand in (\ref{GFZUFBV}) for
$J_A=0$ is invariant under the following global transformations:
\beq
\label{BRSTBV}
\delta\phi^A = \lambda^A\mu
\,,\qquad
\delta\phi^*_A = \mu\frac{\delta S}{\delta\phi^A}
\,,\qquad
\delta\lambda^A = 0\,.
\eeq
It is very important to remember that the existence of this 
symmetry follows from the fact that the bosonic functional 
$S$ satisfies the generating equation (\ref{MastEBV}). 
The transformations (\ref{BRSTBV}) represent the 
BRST-transformations in  the space of variables 
$\phi,\;\phi^*,\;\lambda$.

The symmetry of the vacuum functional $Z(0)$ under the BRST
transformations (\ref{BRSTBV}) paves the way for establishing
an independence of the $S$ matrix on the choice of gauge in the
BV-quantization. Indeed, suppose $Z_{\Psi}\equiv Z(0)$. We shall 
change the gauge $\Psi\rightarrow\Psi + \delta\Psi$. In the 
functional integral for $Z_{\Psi+\delta\Psi}$ we make the 
change of variables, choosing for $\mu$
\beq
\label{mu 1}
\mu = -\frac{i}{\hbar}\delta\Psi.
\nonumber
\eeq
After simple algebraic calculations we find that
\beq
\label{GIZBV}
Z_{\Psi + \delta\Psi} = Z_{\Psi}.
\eeq

In order to derive the Ward identity corresponding to the
BRST-symmetry, it is convenient to consider the extended 
generating functional of the Green functions 
\beq 
\label{GFZExtend} 
{\cal Z}(J,\phi^*)
= \int d{\phi}
\exp \Big\{ \frac{i}{\hbar} \big[S_{\psi}(\phi,\phi^*)
+ J_A\phi^A\big]\Big\},
\eeq
where
\beq
\label{ExtActBV}
S_{\psi}(\phi, \phi^*)
= S\Big( \phi,\,\phi^* + \frac{\de\Psi}{\de}\phi \Big).
\eeq
From the above definition it follows that
\beq
\label{mu 2}
{\cal Z}(J,\phi^*)|_{\phi^* = 0}= Z(J),
\nonumber
\eeq
where $Z(J)$ has been defined in (\ref{ZBV}).
\ From BRST symmetry follows the Ward identity for the extended
generating functional of the Green's functions
\beq
\label{WIZBV}
J_A\frac{\delta {\cal Z}}{\delta\phi^*_A} = 0\,.
\eeq

Introducing the generating functional of connected Green's 
functions,
\ ${\cal W}={\cal W}(J, \phi^*) = -i\hbar \ln {\cal Z}$, \
the identity (\ref{WIZBV}) can be rewritten as
\beq
\label{WIWBV}
J_A\frac{\delta {\cal W}}{\delta\phi^*_A} = 0.
\eeq

The generating functional of the vertex functions
 (Effective Action)
\ $\Gamma = \Gamma (\phi,\;\phi^*)$ \ is introduced in a standard
way, through the Legendre transformation of ${\cal W}$,
\beq
\label{EA}
\Gamma (\phi,\;\phi^*) = {\cal W}(J,\phi^*) - J_A\phi^A
\,,\qquad
\phi^A = \frac{\delta{\cal W}}{\delta J_A}
\,,\qquad
\frac{\delta\Gamma}{\delta\phi^A}=-J_A\,.
\eeq
Finally, the Ward identity for the generating functional
of the vertex functions can be obtained directly from
(\ref{WIWBV}) and (\ref{EA}), in the form
\beq
\label{WIGammaBV}
(\Gamma,\Gamma) = 0\,.
\eeq
The Ward identity (\ref{WIGammaBV}) has universal form and 
plays a very important role in proof of gauge invariant renormalizability of general gauge theories \cite{VLT}. 
In deriving this identity all fields under consideration 
have been assumed to be quantized. However, it looks evident 
that the form of Eq. (\ref{WIGammaBV}) will be the same 
in presence of external background (for example, a 
gravitational background) fields as well (see below).
In the next section we will see that this equation 
represents a suitable basis for the consideration of 
quantum field theory in curved space.

\section{General gauge theories in curved space}

Let us consider a theory of gauge fields $A^i$ in an
external gravitational field $g_{\mu\nu}$.
The classical theory is described by the action which
depends on both dynamical fields and external metric,
\beq
 S_0=S_0(A,g)\,.
\label{action}
\eeq
Here and below we use the condensed notation
$g\equiv g_{\mu\nu}$ for the metric, when it is an
argument of some functional or function. The action
(\ref{action}) is assumed to be gauge invariant,
\beq
 S_{0,i}R^i_a=0,\quad \delta A^i=R^i_a(A,g)\lambda^a
\,,\quad \lambda^a = \lambda^a(x)\quad (a=1,2,...,n)\,,
\label{gauge} \eeq as well as covariant, \beq \delta_g S_0 &=&
\frac{\delta S_0}{\delta A^i}\delta_g A^i + \frac{\delta S_0}{\delta
g_{\mu\nu}}\delta_g g_{\mu\nu}=0\,, \label{diff} \eeq where
$\lambda^a$ are independent parameters of the gauge transformation,
corresponding to the symmetry group of the theory. The
diffeomorphism transformation of the metric in Eq. (\ref{diff}) has
the form \beq \delta_g g_{\mu\nu} &=& -
g_{\mu\alpha}\partial_{\nu}\xi^{\alpha} -
g_{\nu\alpha}\partial_{\mu}\xi^{\alpha} -
\partial_{\alpha}g_{\mu\nu}\xi^{\alpha} \nonumber
\\
&=& - g_{\mu\alpha}\nabla_{\nu}\xi^{\alpha}
- g_{\nu\alpha}\nabla_{\mu}\xi^{\alpha}
\,=\, -\nabla_{\mu}\xi_{\nu}-\nabla_{\nu}\xi_{\mu}\,.
\label{xi}
\eeq
Here $\xi^\al$ are the parameters of the coordinates
transformation,
\beq
&& \xi^{\alpha} = \xi^{\alpha}(x)
\quad (\alpha=1,2,...,d)\,.
\eeq

As usual, an
explicit expression for $\delta_g A^i$ depends on tensor
(or spinor) properties of $A^i$. For example, in the
case of a scalar field $A$ one has
$\delta_g A=-\partial_{\alpha}A\xi^{\alpha}$ while
in the case of a vector field $A^{\mu}$ the transformation
rule is
$\,\delta_g A^{\mu} = A^{\nu}\nabla_{\nu}\xi^{\mu}
- \xi^{\nu}\nabla_{\nu}A^{\mu}$, \ etc.
In general, our interest is to explore the renormalization
properties of the theories which include all three kind of fields
(fermions, vectors and scalars), such that, for instance, the
Standard Model and its extensions, including Grand Unified 
Theories (GUTs), would be covered. Therefore the notation 
$A^i$ in (\ref{gauge}) and (\ref{diff}) means the set of 
fields with the different transformation rules.

The generating functional $Z(J,\phi^*,g)$ of the Green functions 
can be constructed in the form of the functional integral 
\beq 
\label{Z}
{\cal Z}(J,\phi^*,g)=\int d{\phi} \exp\Big\{\frac{i}{\hbar}
\Big[S_{\psi}(\phi,\phi^*,g) +J_A{\phi}^A\Big]\Big\}. 
\eeq
Here\footnote{We restrict ourself to the case of irreducible 
close gauge theories only, in order to simplify the 
description of the configuration space.}
$\phi^A=(A^i,B^a,C^a, {\bar C}^a)$ represents the full set of 
fields of the complete configuration space of the theory under
consideration and 
$\phi^*_A=(A^*_i,B^*_a,C^*_a, {\bar C}^*_a)$ are
corresponding antifields. Finally,
$S_{\psi}(\phi,\phi^*,g)$
is the quantum action constructed with the help of
the solution \ $S=S(\phi,\phi^*,g)$ \ of the master equation
\beq
\label{ME}
(S,S) = 0\,, \qquad
S(\phi,\phi^*,g)|_{\phi^*=0} = S_0(A,g)
\eeq
in the form
\beq
S_{\psi}(\phi,\phi^*,g)
\,=\, S\big(\phi,\phi^*
+ \frac{\delta\Psi(\phi,g)}{\delta \phi},g\big)\,.
\label{Sqa}
\eeq
In the last equation, (\ref{Sqa}),
$\Psi(\phi,g)$ is a gauge fixing functional.
Note that $S_{\psi}$ satisfies the master equation
\beq
 (S_{\psi},S_{\psi})=0.
\eeq

From the gauge invariance of initial action (\ref{gauge}),
in the usual manner one can derive the BRST symmetry and
the Ward identities for generating functionals
${\cal Z},{\cal W}$ and $\Gamma$ in the form (\ref{WIZBV}), 
(\ref{WIWBV}) and (\ref{WIGammaBV}) respectively.

A solution to the master equation (\ref{ME}) can be always
found in form of a series in antifields $\phi^*$
(see \cite{BV}),
\beq
\label{exp}
S(\phi,\phi^*,g)
\,=\,S_0(A,g)+A^*_iR^i_a(A,g)C^a+{\bar C}^*_aB^a+\cdots\,,
\eeq
where dots mean higher order terms in fields $B^a, C^a$.
We assume that every term in (\ref{exp}) is transformed as
a scalar under arbitrary local transformations of
coordinates $x^{\mu}\rightarrow x^{\mu}+\xi^{\mu}(x)$.
It means the general covariance of $S=S(\phi,\phi^*,g)$,
\beq
\label{cov}
\delta_g S(\phi,\phi^*,g)
= \frac{\delta S}{\delta \phi^A}\delta_g \phi^A
+ \delta_g \phi^*_A\frac{\delta S}{\delta \phi^*_A}
+ \frac{\delta S}{\delta g_{\mu\nu}}\delta_g g_{\mu\nu}
= 0.
\eeq

Let us choose the gauge fixing functional
$\Psi=\psi(\phi,g)$ in a covariant form
\beq
\delta_g \Psi=0\,,
\eeq
then the quantum action $S_{\psi}=S_{\psi}(\phi,\phi^*,g)$
obeys the general covariance too
\beq
\label{covqa}
\delta_g S_{\psi}=0\,.
\eeq

From the Eq. (\ref{covqa}) and the assumption that the term
with the sources $J_A$ in (\ref{Z}) is covariant
\beq
\label{covJ}
\delta_g (J_A\phi^A)
= (\delta_g J_A)\phi^A+J_A(\delta_g\phi^A) = 0\,,
\eeq
follows the general covariance of ${\cal Z}={\cal Z}(J,\phi^*,g)$.
Indeed,
\beq
\label{covZ}
\delta_g {\cal Z}(J,\phi^*,g)
&=&
\frac{i}{\hbar}\int d{ \phi}
\Big[
\de_g \phi^*_A\frac{\de S_{\psi}({\phi},\phi^*,g)}
{\delta \phi^*_A}
+ \frac{\delta S_{\psi}({\phi},\phi^*,g)}
{\delta g_{\mu\nu}}\delta_g g_{\mu\nu}+
(\delta_g J_A){\phi}^A\Big] 
\\
\nonumber
&\times&
\exp\Big\{\frac{i}{\hbar}\Big[
S_{\psi}({\phi},\phi^*,g)
+ J_A{\phi}^A\Big]\Big\}\,.
\eeq

Making change of integration variables in the
functional integral, (\ref{covZ}),
\beq
{\phi}^A\;\;\;\rightarrow\;\;\;\phi^A+\delta_g\phi^A\,,
\eeq
we arrive at the relation
\beq
\nonumber
&&
\delta_g {\cal Z}(J,\phi^*,g)
= \frac{i}{\hbar}\int d\Phi
\Big[\frac{\delta S_{\psi}}{\delta \phi^A}\delta_g \phi^A
+ \delta_g \phi^*_A\frac{\delta S_{\psi}}{\delta \phi^*_A}
\\
&+& \frac{\delta S_{\psi}}{\delta g_{\mu\nu}}\delta_g g_{\mu\nu}
+ (\delta_g J_A)\phi^A+J_A(\delta_g\phi^A)\Big]
\,\exp\Big\{\frac{i}{\hbar}\Big[S_{\psi}(\phi,\phi^*,g)
+ J_A\phi^A\Big]\Big\}
\\
&=&
\frac{i}{\hbar}\int d\phi \Big[\delta_g S_{\psi}
+ \delta_g (J_A\phi^A)\Big]
\,\exp\Big\{\frac{i}{\hbar} \Big[S_{\psi}(\phi,\phi^*,g)
+ J_A\phi^A\Big]\Big\}\,=\,0\,
\label{covZ1}.
\eeq
From (\ref{covZ1}) follows that the generating
functional of connected Green functions ${\cal W}(J,\phi^*,g))$
\beq
{\cal W}(J,\phi^*,g)=\frac{i}{\hbar}\ln Z(J,\phi^*,g)
\eeq
obeys the property of the general covariance as well
\beq
\label{covW}
\delta_g {\cal W}(J,\phi^*,g)=0\,.
\eeq

Consider now the generating functional of vertex functions
$\Gamma=\Gamma(\phi,\phi^*,g)$
\beq
\Gamma(\phi,\phi^*,g)= {\cal W}(J,\phi^*,g)-J_A\phi^A\,,
\eeq
where
\beq
\label{Gamma}
\phi^A=\frac{\delta {\cal W}(J,\phi^*,g)}{\delta J_A},\quad J_A=-
\frac{\delta \Gamma(\phi,\phi^*,g)}{\delta \phi^A}.
\eeq
From definition of $\phi^A$ (\ref{Gamma}) and the general covariance of
$W(J,\phi^*,g)$
we can conclude the general covariance of $J_A\phi^A $.
Therefore,
\beq
\label{covGamma}
\delta_g\Gamma(\phi,\phi^*,g)= \delta_g {\cal W}(J,\phi^*,g)=0.
\eeq

\section{Gauge-invariant renormalization in curved space-time}

Up to now we have considered non-renormalized generating 
functionals of Green functions. 
The next step is to prove the general covariance for
renormalized generating functionals. For this end, let us 
first consider the one-loop approximation for \
$\Gamma=\Gamma(\phi,\phi^*,g)$, 
\beq 
\label{Gamma1loop} 
\Gamma = S_{\psi} + {\bar \Ga}^{(1)} 
= S_{\psi} + \hbar \big[{\bar \Gamma}^{(1)}_{div}
+ {\bar\Gamma}^{(1)}_{fin}\big] + O(\hbar^2)\,,
\eeq 
where ${\bar\Gamma}^{(1)}_{div}$ and ${\bar\Gamma}^{(1)}_{fin}$
denote the divergent and finite parts of the one-loop approximation
for $\Gamma$. The divergent local\footnote{The discussion of
locality of the divergent part of effective action will be given in
the next section.} term ${\bar\Gamma}^{(1)}_{div}$ gives the first
counterpart in the one-loop renormalized action $S_{\psi 1}$ 
\beq
\label{S1loop} 
S_{\psi}\rightarrow S_{\psi 1} =
S_{\psi} - \hbar{\bar\Gamma}^{(1)}_{div}. 
\eeq 
From (\ref{covqa}) and
(\ref{covGamma}) follows that in one-loop approximation we have
\beq 
\label{covGamma1loop} 
\delta_g \big[{\bar\Gamma}^{(1)}_{div}
+ {\bar\Gamma}^{(1)}_{fin}\big] = 0 
\eeq
and therefore ${\bar\Gamma}^{(1)}_{div}$ and
${\bar\Gamma}^{(1)}_{fin}$ obey the general covariance 
independently
\beq 
\label{covGamma1loops} \delta_g{\bar\Gamma}^{(1)}_{div}=0
\,,\qquad \delta_g{\bar\Gamma}^{(1)}_{fin}=0\,. 
\eeq

In its turn the one-loop renormalized action $S_{\psi 1}$ 
(i.e., classical action, renormalized at the one-loop level)
is covariant
\beq
\label{covS1loop}
\delta_g S_{\psi 1}=0\,.
\eeq
Constructing the generating functional of one-loop
renormalized Green functions ${\cal Z}_1(J,\phi^*,g)$, with the
action $S_{\psi 1}=S_{\psi 1}(\phi,\phi^*,g)$, and
repeating  arguments given above, we arrive at the
relation
\beq
\label{covF1loops}
\delta_g {\cal Z}_1=0
\,,\qquad
\delta_g W_1 = 0
\,,\qquad
\delta_g \Gamma_1=0\,.
\eeq
In the last equation we have introduced the new
useful notation for the renormalized up to the one-loop 
order effective action $\Ga_1$. This functional 
includes the contributions of one-loop and also 
higher loop orders, however, only the one-loop 
divergences are removed by renormalization. This 
means that $\Ga_1$ is finite in the ${\cal O}(\hbar)$
order, but may be divergent starting from 
${\cal O}(\hbar^2)$ and beyond. 

The generating functional of vertex functions 
\ $\Ga_1 = \Ga_1(\phi,\phi^*,g)$ \ which
is finite in the one-loop approximation, can be 
presented in the form
\beq
\label{Gamma2loop}
\Ga_1 = S_{\psi}+ \hbar{\bar\Gamma}^{(1)}_{fin}
+ \hbar^2
\big[{\bar\Gamma}^{(2)}_{1,div} 
+ {\bar\Gamma}^{(2)}_{1,fin}\big] + O(\hbar^3)\,.
\eeq
Indeed, this functional contains a divergent part 
${\bar\Gamma}^{(2)}_{1,div}$ and defines renormalization 
of the action $S_{\psi}$ in the two-loop approximation
\beq
\label{S2loop}
S_{\psi}\rightarrow S_{\psi 2} = S_{\psi 1}
- \hbar^2{\bar\Gamma}^{(2)}_{1,div}\,.
\eeq
Starting from (\ref{covGamma1loops}), (\ref{covS1loop})
and  (\ref{covF1loops}) we derive
\beq
\label{covGamma2loops}
\delta_g {\bar\Gamma}^{(2)}_{1,div}
\,=\, 0
\,,\qquad
\delta_g {\bar\Gamma}^{(2)}_{1,fin}=0\,.
\eeq
The last equation means that the general covariance 
condition is satisfied separately for the divergent and 
finite parts of ${\bar\Gamma}_1$ in the two-loop 
approximation. As a consequence, the two-loop renormalized 
action $S_{\psi 2} = S_{\psi 2}(\phi,\phi^*,g)$ is a 
covariant functional
\beq
\label{covS2loop}
\delta_g S_{\psi 2}=0.
\eeq
\vskip 1mm

Applying the induction method we can repeat the procedure
to an arbitrary order of the loop expansion. In this way
we arrive at the followings results:
\vskip 1mm

{\bf a)} \ The full renormalized action,
$S_{\psi R}=S_{\psi R}(\phi,\phi^*,g)$,
\beq
\label{SRloop}
 S_{\psi R}\,=\,
S_{\psi}-\sum_{n=1}^{\infty}\hbar^n {\bar\Gamma}^{(n)}_{n-1,div}\,,
\eeq
which is local in each finite order in $\hbar$, obeys the
general covariance
\beq
\label{covSRloop}
\delta_g S_{\psi R}\,=\,0\,;
\eeq
\vskip 1mm

\vskip 1mm

{\bf b)} \  The renormalized generating functional of vertex
functions, $\Gamma_R=\Gamma_R(\Phi,\Phi^*,g))$, 
\beq
\label{GammaRloop} 
\Ga_R\,=\,
S_{\psi} + \sum_{n=1}^{\infty}\hbar^n {\bar\Gamma}^{(n)}_{n-1,fin}
\,, 
\eeq 
which is finite in each finite order in $\hbar$, is covariant 
\beq
\label{covGammaRloop} 
\delta_g \Gamma_R\,=\,0\,. 
\eeq

It was proved in \cite{VLT} that the renormalized action
$S_{\psi R}$ satisfies the master equation 
\beq 
\label{MESRloop}
(S_{\psi R},S_{\psi R})=0 
\eeq 
and the Ward identities for non-renormalized and 
renormalized generating functionals of vertex
functions have the form 
\beq 
\label{MEGammaRloop}
(\Gamma,\Gamma)\,=\,0\,,\qquad (\Gamma_R,\Gamma_R)\,=\,0\,. 
\eeq

The last equations mean that the gauge invariant 
renormalizability (\ref{MEGammaRloop}) of a quantum 
field theory takes place in the presence of an external 
gravitational field, such that the general covariance 
of Effective Action (\ref{covGammaRloop}) is also 
preserved. In order to use this important result we 
have to perform an additional consideration and check 
how the covariance is preserved in case when we use 
apparently non-covariant techniques, e.g., related to 
the representation of the metric as a sum of the flat 
one and perturbation. This subject will be treated in 
the next section. 

\section{Non-covariant gauges}

In many cases it is interesting to consider the renormalization 
of quantum field theory in curved space using the non-covariant 
gauge fixing functionals. One important example of such 
consideration can be found in Sect. 7 of the present article, 
where we discuss power counting renormalizability in curved 
space. Let us see how the non-covariant gauge fixing can be 
implemented in the quantum theory. 

Our purpose is to investigate the problem of general covariant
renormalizability for general gauge theories in the presence of 
an external gravitational field, when one uses non-covariant 
gauge fixing functional $\Psi=\Psi(\phi,g)$, 
\beq
\label{ncovG}
\delta_g \Psi\neq 0\,.
\eeq
As before, we assume that the classical action of the theory 
$S=S(\phi,\phi^*,g)$  is covariant, i.e. $\delta_gS=0$, but now the action
$S_{\psi}=S_{\psi}(\phi,\phi^*,g)=S(\phi,\phi^*+\delta\Psi/\delta\phi,g)$ 
is not covariant, $\de_g S_{\psi}\neq 0$.
Our consideration will be essentially based on the known formalism 
for investigating the gauge dependence in general gauge theories, 
given in \cite{VLT}. Non-covariance of $S_{\psi}$ can be described 
in the form of anticanonical infinitesimal 
transformation with the odd generating functional
\beq
 && X(\phi,\phi^*,g)=\phi^*_A\phi^A+\delta_g\Psi(\phi,g)\,,
\\
\nonumber 
\\
\Phi^A &=& \frac{\delta X(\phi',\phi^*,g)}{\delta\phi^*_A}
= \Phi^{A'},\quad \phi^{*'}_A
= \frac{\delta X(\phi',\phi^*,g)}{\delta\phi^{'}_A} 
= \phi^*_A+\frac{\delta\delta_g \Psi}{\delta\phi^A}\,,
\eeq
when 
\beq
\label{ncovAc}
 \delta_g S_{\psi}
= \frac{\delta\delta_g \Psi}{\delta\phi^A}
 \frac{\delta S_{\psi}}{\delta\phi^*_A}
= (\delta_g \Psi, S_{\psi})\,.
\eeq

The variation of $S_{\psi}$ (\ref{ncovAc}) leads to
the variations of generating  functionals of the 
Green functions
${\cal Z}={\cal Z}(J,\phi^*,g)$, connected Green functions 
${\cal W}={\cal W}(J,\phi^*,g)$ and vertex functions
$\Gamma=\Gamma(\phi,\phi^*,g)$ in the form
\beq
\label{ncovZ}
 \delta_g {\cal Z}
&=& \frac{i}{\hbar}J_A\frac{\delta}{\delta\phi^*_A}
\delta_g \Psi \Big(\frac{\hbar}{i}
\frac{\delta}{\delta J},g\Big){\cal Z}\,,
\\
\label{ncovW} 
\delta_g {\cal W}
&=& J_A\frac{\delta}{\delta\phi^*_A}
\langle\delta_g \Psi\rangle\,,
\\
\label{ncovGamma} 
\delta_g\Gamma
&=&
(\langle\langle \delta_g \Psi\rangle\rangle,\Gamma)\,,
\eeq
where the notations
\beq
\langle\delta_g \Psi\rangle
&=& \delta_g \Psi\Big(\frac{\delta {\cal W}}{\delta J}
+ \frac{\hbar}{i}\frac{\delta}{\delta J},g\Big)\,, 
\nonumber
\\
\langle\langle \delta_g \Psi\rangle\rangle 
&=& \delta_g \Psi\Big(\phi+i\hbar(\Gamma^{''})^{-1} 
\frac{\delta_l}{\delta \phi},g\Big)\,,
\nonumber
\\
 \Gamma^{''}_{AB} &=& \frac{\delta_l}{\delta
\phi^A}\frac{\delta}{\delta \phi^B}\Gamma.
\eeq
were used. These results can be immediately reproduced 
in the renormalized theory \cite{VLT}. Namely, for 
the variation (\ref{ncovAc}), the corresponding variation 
of renormalized action $\delta_g S_{\psi R}$ can be 
presented in the form
\beq
\delta_g  S_{\psi R}=(\delta_g \Psi_R,S_{\psi R})
\eeq
of the anticanonical transformation with local generating 
functional \ $X=\phi^*_A\phi^A+\delta_g \Psi_R$, 
\beq
\delta_g\Psi_R(\phi,\phi^*,g)=\delta_g\Psi(\phi,g)
- \sum^{\infty}_{n=1}
\hbar^n\delta_g\Psi^{(n)}_{n-1,div}(\phi,\phi^*,g)\,,
\eeq
while the variation of renormalized vertex generating functional
$\delta_g\Gamma_R$ has the form
\beq
\delta_g\Gamma_R=(\langle\langle
\delta_g\Psi_R\rangle\rangle_R,\Gamma_R),
\label{trans1}
\eeq
which corresponds to finite anticanonical transformation  with
generating function
\beq
X=\phi^*_A\phi^A+\langle\langle \delta_g
\Psi_R\rangle\rangle_R,\quad\langle\langle
\delta_g\Psi_R\rangle\rangle_R= \delta_g\Psi(\phi,g)
+ \sum^{\infty}_{n=1}\hbar^n\delta_g\Psi^{(n)}_{n-1, fin}.
\label{trans2}
\eeq
\\
In the formulas presented above 
we have used the notations $\delta_g\Psi^{(n)}_{n-1, div}$ 
and $\delta_g\Psi^{(n)}_{n-1, fin}$ for the  divergent and 
finite terms, respectively, of the n-loop approximation 
for the generating function of an anticanonical 
transformation which is finite in $(n-1)$-th order 
approximation and is constructed on the basis of the 
theory with the action $S_{\psi(n-1)}$.

The interpretation of the relations (\ref{trans1}) 
and  (\ref{trans2}) is that the theory with external
gravitational field may have non-covariance in the 
renormalized effective action, but it comes 
only from the possible non-covariance of the arguments. 
Here the expression arguments is used to denote the 
full set of the mean fields from which the effective 
action depends, as defined in (\ref{EA}). 
\
Therefore, the violation of the general coordinate 
symmetry which can occur because of the non-covariant 
gauge-fixing can be always included into the arguments. 
As a consequence, one can always define some special 
set of arguments, in terms of which the quantum dynamics 
is decribed in a completely covariant way. One impotant 
aspect of this feature is that we can actually perform 
general considerations or make practical calculations 
in a non-covariant gauges. After that we can always 
restore the covariance using those parts of effective 
action which are not affected by gauge transformation. 
A practical examples of this tecnique can be  
found in many publications, but here we constructed 
a theoretical background for its consistent us. 
In the next sections we will see, also, that this 
result opens the way for a practical construction 
of renormalizable gauge theories in curved space-time. 

 Note that there exists 
another interpretation of the gauge dependence of effective action (see \cite{L}).
Namely it can be proved that dependence on the gauge of effective action is 
proportional to its extremals, i.e. physical quantities calculeted on shell do not 
depend on the gauge.

\section{On the locality of the counterterms}

In most cases the general consideration of renormalizability
is based on the hypothesis of locality of all necessary
counterterms. This statement was first proved in general
form is \cite{Weinberg-1960} and is known as Weinberg
theorem. One can find a more pedagogical consideration of
this theorem in the book \cite{Collins}.
It is important for us to understand whether the locality
of the counterterms holds for the case when the external
gravitational field is present. It is easy to see that the
arguments of  \cite{Collins} can be taken carefully in this
case and, in principle, some special attention to this
issue is in order. Here we present a qualitative
consideration which shows that the locality of the
counterterms still holds in the presence of external gravity.

Let us consider the theory of the matter fields $A\equiv A^i$ with
the action (\ref{action}), which depends also on the external metric
$g\equiv g_{\mu\nu}$, \ $S_0(A,g)$. In order to discuss the locality
of the counterterms it proves useful to parameterize the metric as
\beq g_{\mu\nu} = \eta_{\mu\nu} + h_{\mu\nu}\,, \label{flat} \eeq
where we do not need to make special assumptions about the field
$h_{\mu\nu}$. Starting from the parametrization (\ref{flat}) of the
metric one can construct the diagrammatic representation of the path
integral (\ref{Z}). The relevant Feynman diagrams include external
lines of the fields ${\tilde\Phi}$ only, and the external lines of
both quantum fields (given by sources in the Schwinger formalism)
and the classical background field $h_{\mu\nu}$.

How can we know that the presence of the background field
$h_{\mu\nu}$ does not lead to the nonlocal counterterms
at higher orders of the loop expansion? In order to address
this question, let us consider the quantum gravity completion
of the theory. This means we start from the extended
classical action
\beq
S_0^{ext} = S_0(A,g) + S_{QG}\,,
\label{QGcomplete}
\eeq
where $S_{QG}$ is an action of a quantum gravitational
field. As far as we do not care about power counting
renormalizability of the theory at this stage (see the
next section for the corresponding discussion), $S_{QG}$
can be just the Einstein-Hilbert action. Another possibility
is to include the higher derivative terms. In fact, as
we shall see in a moment, the result does not depend on
the choice of the action  $S_{QG}$. Let us also remark
that the path integral representation of the quantum
gravitational theory includes also a set of ghost and
antighost fields (see, e.g., \cite{Stelle,VorTyu84,book}
for the higher derivative case). For the sake of simplicity
we will not write these extra fields here, or assume they
are included automatically into $\Phi^{*}$.

One can note that the new theory, based on the action
(\ref{QGcomplete}), includes internal lines of the metric
field $h_{\mu\nu}$ and does not include external fields.
Therefore the Weinberg theorem can be applied and we
can use the result for the locality of the counterterms
at any loop order in the complete theory. In particular,
one can prove that only local solutions of the master
equations can be relevant for the divergences in the case
of the fourth derivative quantum gravity \cite{VorTyu84}.
Moreover the proof presented in \cite{VorTyu84} does not
require the details of the action of quantum gravity and
indeed can be generalized for other cases, including the
quantum General Relativity.

On the other hand, the theory with the quantum gravity
completion includes all those Feynman diagrams which
give contribution to divergences of the theory with
external metric. Therefore, since the complete theory
does not have nonlocal divergences, the reduced one
with external metric does not have them either.
Hence, for the usual quantum field theory on curved
background we have strong reasons to assume the 
locality of the necessary counterterms, to all orders
in the loop expansion.

One more observation is in order. All arguments
presented above correspond to the usual quantum field
theory on curved background and can be violated in
the case we consider the theory with spontaneous
symmetry breaking \cite{sponta}. In this case the
nonlocalities show up already at the classical level,
in the induced action of gravity. At the quantum
level, the non-local structures get renormalized and
hence we are forced to introduce an infinite set of
non-local counterterms. However, the details of the
consideration presented in \cite{sponta} show that
the mentioned non-localities are always related to
the scalar (Higgs) field, such that the corresponding
renormalization becomes local if this field is treated
as an independent one.

\section{Power-counting renormalizability and construction
of renormalizable theories}

In the previous sections we have shown that the 
non-anomalous gauge theory in curved space-time is 
renormalizable in a sense that the necessary counterterms, 
in all orders of the loop expansion, are given by the local, 
covariant and gauge invariant expressions. This fact enables 
one to prepare the receipt of constructing the renormalizable 
theories in curved space.

Let us consider the \ $h_{\mu\nu}=g_{\mu\nu}-\eta_{\mu\nu}$
\ parametrization of the external metric, which enables one
to deal with the usual flat-space Feynman diagrams. Compared
to the diagrams of the flat space-time theory these diagrams
have external lines of the metric field \ $h_{\mu\nu}$.
As far as gravity is non-polynomial interaction, there may
be, in principle, unrestricted amount of such external
lines coming to any vertex of the diagram. However, the
covariance of the counterterms which we have proven in
Sect. 5, enables one to establish the general form of 
the counterterms.

We start from the case of a scalar field  \ $\ph$ \ with
the \ $\la\ph^4$-interaction. The first diagram we will be
interested in is the one-loop correction vertex function.
The situation which occurs in curved space-time
is illustrated in the Fig. 1. One can note that the
lines of the field \ $h_{\mu\nu}$ \ may either
produce new vertices or be connected to the existing
vertex due to the expansion
\beq
\sqrt{- g}\la\ph^4
= \la\ph^4\cdot \Big[ 1 + \frac12\,h
+\frac18\,h^2 - \frac14\,h_{\mu\nu}h^{\mu\nu}
+ \,...\Big]\,\,,
\qquad
h = h_{\mu\nu}g^{\mu\nu}\,.
\label{action 6}
\eeq
It is easy to see that the first kind of diagrams
has more propagators in the loop that the initial
flat-space diagram. The typical examples are the
diagrams in the second line in Fig. 1. It is obvious
that the divergence of the diagrams with larger number
of propagators will be smaller. For instance, the
mentioned diagrams in the second line are all finite.
On the other hand the diagrams with the lines
of  \ $h_{\mu\nu}$ \ connected only to the vertices
will sum up to produce the logarithmic divergences
which will be exactly of the form of the flat-space
divergence, multiplied by the $\sqrt{-g}$, defined
in (\ref{action 6}). Any
other form would enter in conflict with locality
and covariance of the divergences which we have
proven in the previous sections\footnote{The
explicit calculations in the momentum-subtraction
scheme confirm this conclusion \cite{bexi}. Also
they show that the finite part of the vertex
function is a nonlocal object, as it usually
happens. Of course, this does not contradict the
Weinberg theorem \cite{Weinberg-1960,Collins} which
concerns only the UV divergences.}.
\vskip 8mm

 \begin{tabular}{c}
 \mbox{\hspace{+1.0cm}}
 \includegraphics[width=12cm,angle=0]{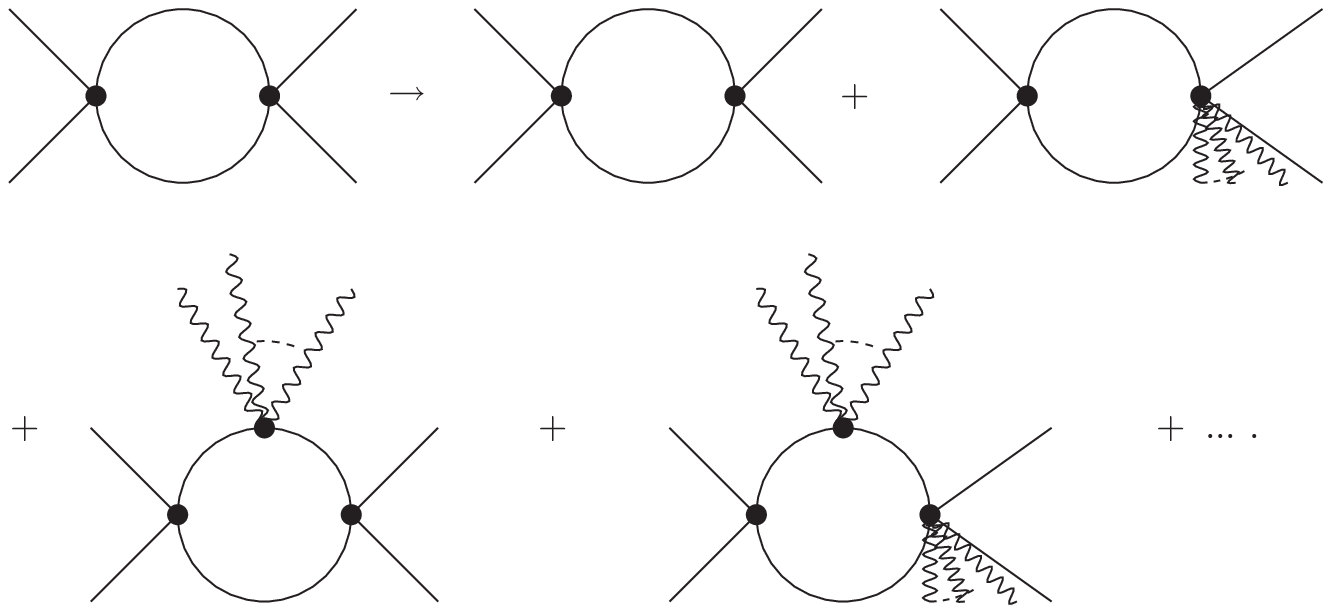}
 \end{tabular}
\begin{quotation}
{\bf Figure 1.} \ The single diagram with quadratic
divergences in flat space generates an infinite set
of diagrams with external lines of $h_{\mu\nu}$. Some
of those diagrams have quadratic or logarithmic
divergences, others are finite.
\end{quotation}

As the next step let us consider the one-loop contribution
to the field propagator, which has quadratic divergence
in flat space-time case. The situation which occurs in
curved space-time is illustrated in the Fig. 2. Again,
as in the case of the vertex diagram, one can distinguish
the two kinds of diagrams. The first kind of diagrams
has more propagators in the loop, compared to the initial
flat-space diagram. The typical examples are the last
diagram in the first line and the last two diagrams
in the second line on Fig. 2. It is obvious that
the divergence of the diagrams with larger number
of propagators will be smaller. For instance, the
initial flat-space diagram on Fig. 2 has quadratic
divergences and the last diagram in the first line
has only logarithmic divergences, exactly as all
other diagrams with one extra vertex. Moreover,
the diagrams with two extra vertices are all finite.
\vskip 8mm

 \begin{tabular}{c}
 \mbox{\hspace{+1.0cm}}
 \includegraphics[width=12cm,angle=0]{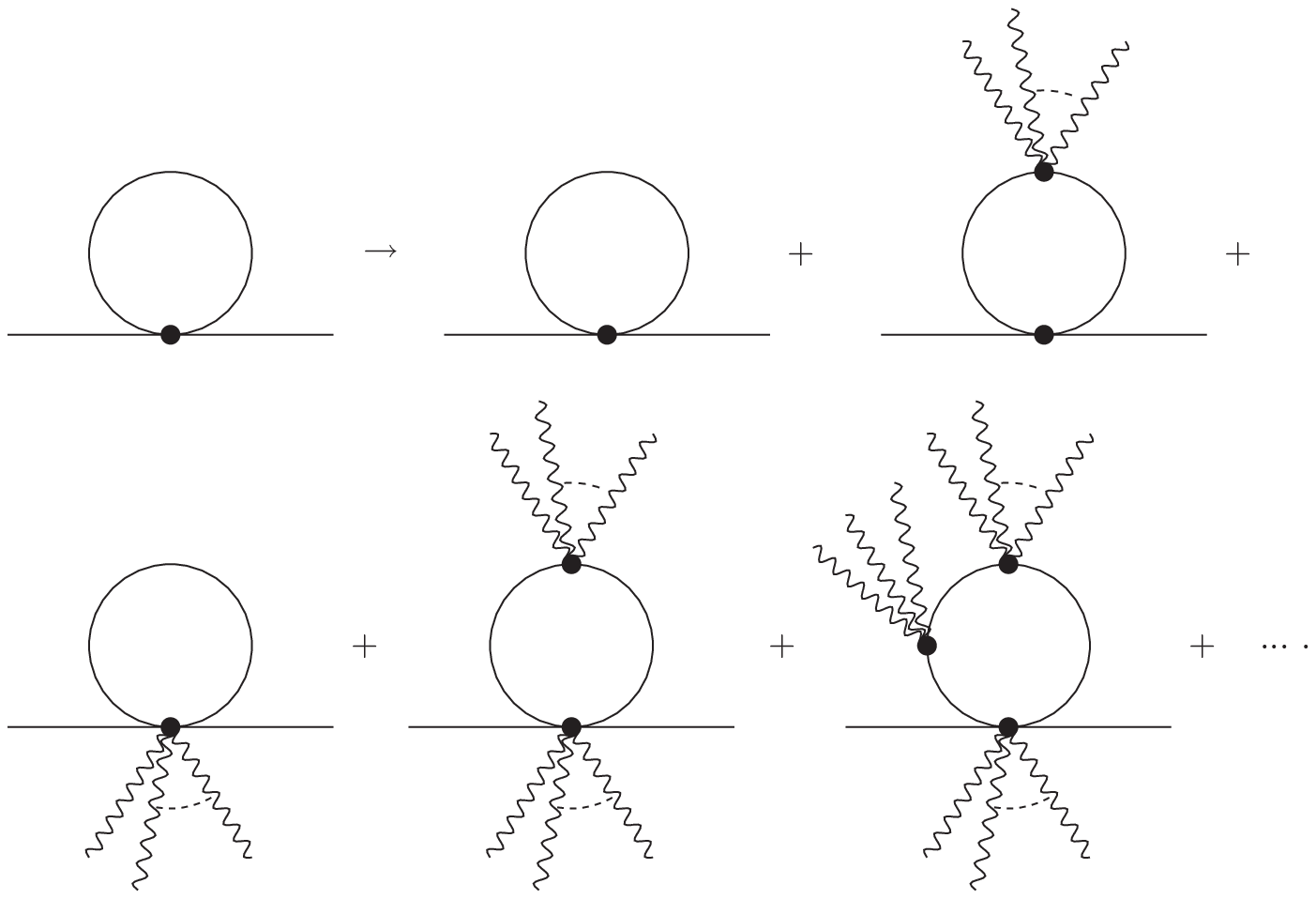}
 \end{tabular}
\begin{quotation}
{\bf Figure 2.} \ The single diagram with quadratic
divergences in flat space generates an infinite set
of diagrams with external lines of $h_{\mu\nu}$. Some
of those diagrams have quadratic or logarithmic
divergences, others are finite.
\end{quotation}

What are the counterterms needed to cancel the new
logarithmic divergences, e.g., the ones produced by
the last diagram in the first line of Fig. 2?
As we already know, this counterterm must be covariant
and local. It is obvious that there can not be
derivatives of the scalar. Furthermore, the
dimensional consideration shows that the correct
dimension of the counterterm can be provided only
by including second derivatives of \ $h_{\mu\nu}$
\ functions. As we know, the only invariant which can
be constructed from the second derivatives of the
metric is the scalar curvature $R$. Therefore the
unique possible form of the counterterm is the
integral of
\beq
\sqrt{-g} R\ph^2\,,
\label{nonminimal}
\eeq
which is called the non-minimal term.

Finally, let us consider the last possible source of the 
one-loop divergences which are the vacuum diagrams. The 
generalization of the single one-loop vacuum diagram in 
flat space to the curved space-time case is demonstrated 
in Fig. 3. It is obvious that the situation is similar to 
the one with the previous diagrams, in a sense that 
inserting the new vertices will produce less divergent
diagrams. The divergences can be classified by a number 
of derivatives of the metric, and we start from the 
zero-derivative case. Both the initial diagram and its 
covariant version have only quartic divergence for the 
massless scalar and, also, quadratic and logarithmic 
divergences in the massive case. All these divergences 
can be removed by renormalizing the covariant 
cosmological constant term 
\ $\int d^4x\sqrt{-g}\rho_\La$, 
which must be, therefore, included into the classical 
action. Let us note that the diagrams corresponding to 
the renormalization of the covariant cosmological 
constant term have only one vertex and no derivatives 
of the external  \ $h_{\mu\nu}$ \ functions.

\vskip 8mm

 \begin{tabular}{c}
 \mbox{\hspace{+1.0cm}}
 \includegraphics[width=12cm,angle=0]{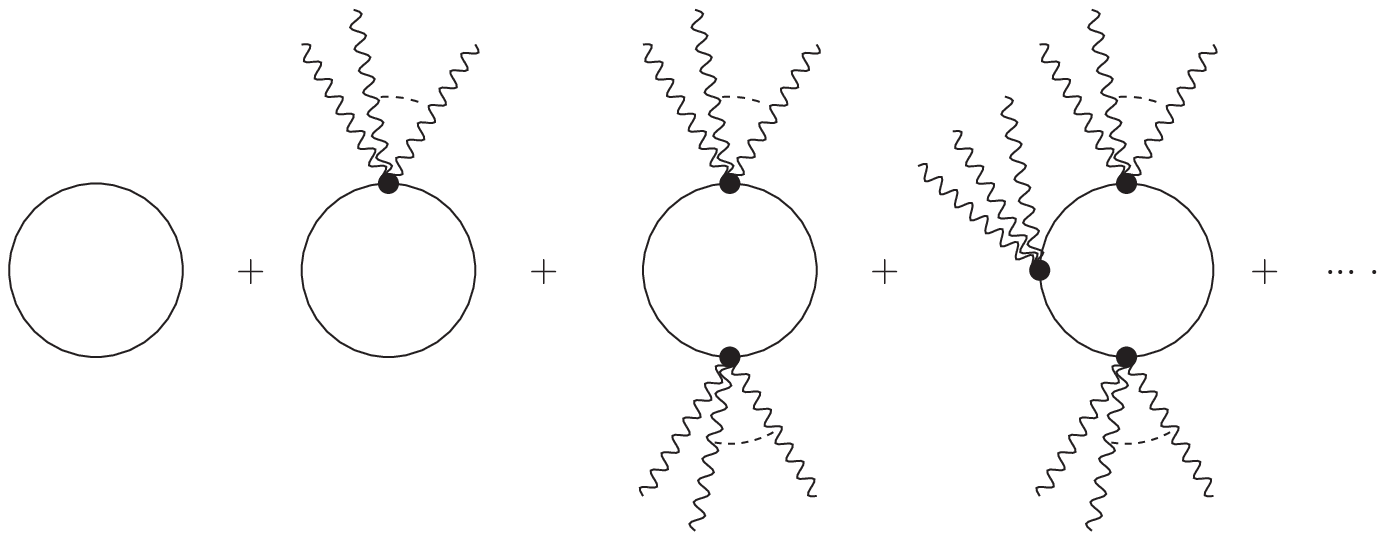}
 \end{tabular}
\begin{quotation}
{\bf Figure 3.} \ The single diagram with quartic 
divergences in flat space leads to the diagrams with 
quartic, quadratic and logarithmic divergences due 
to external lines of $\,h_{\mu\nu}\,$ with the new 
vertices. Despite there are infinitely many new diagrams, 
the divergences are well controlled by covariance.
\end{quotation}

Since the initial diagram has quadratic divergences, the ones with
one new vertex will have quadratic and (in case of massive scalar)
logarithmic divergence. The analysis is pretty much the same as in
the case of the diagrams from Fig. 2. It is obvious, from the
dimensional reasons and covariance, that the quadratic divergence
will be removed by the counterterm linear in curvature and the
logarithmic ones by the counterterm proportional to
\beq 
\it d^4x \sqrt{-g}\, Rm^2\,, 
\label{m2}
\eeq 
where $m$ is the mass of the scalar field. All these counterterms
can be removed by renormalizing the Einstein-Hilbert term, which 
is also (along with the cosmological term) a necessary element of
renormalizable theory in curved space-time.

Finally, there are logarithmically divergent diagrams with two new
vertices and with four derivatives of the external \ $h_{\mu\nu}$ 
\ functions. The covariance and locality show that the necessary
counterterms have the following form 
\beq 
\int d^4x\sqrt{-g} \Big\{
  \al_1 R_{\mu\nu\al\be}R^{\mu\nu\al\be} 
+ \al_2 R_{\mu\nu}R^{\mu\nu}
+ \al_3 R^2 + \al_4 \na^2 R   \Big\}\,. 
\label{HD} 
\eeq 
It is very
important that the possible divergences listed above represent the
complete set and no others can appear. Moreover, this consideration
can be immediately generalized for an arbitrary renormalizable (in
flat space-time) theory including fermions, massless gauge vectors
and scalars. It is easy to see that the counterterms listed above,
plus covariant generalizations of the familiar counterterms in flat
space-time, still represent the complete set. Let us note that the 
non-minimal term is possible only in the scalar sector of the 
theory. According to the consideration performed in Sect. 5 
and Sect. 6, the described structure of divergences is compatible 
with the gauge invariance of the theory at quantum level.

The analysis of the one-loop divergences can be used
to establish the renormalization structure at higher
loops. Let us consider the two-loop divergences. The
one-loop sub-diagrams produce the divergences
described above and can be removed by adding
minimal, nonminimal and vacuum local counterterms. 
As far as these counterterms have the same structure as
the classical action, and the non-local part does not 
influence the second-loop countereterms, the part of the 
one-loop diagrams which is relevant for the divergences
coming from the last integration, is essentially
the same as in flat space, plus non-minimal term. 
Therefore, at the second-loop we meet exactly the same 
types of counterterms as at the one-loop level, which we
havedescribed above. The only difference will be the 
the renormalization coefficients which will have higher 
powers of coupling constants. 

The iteration procedure can be applied to higher
loops and we will always meet the same structure
of renormalization in curved space which was already
described in \cite{book} (see further references
therein). All in all, we can state that the
an arbitrary renormalizable in flat space-time theory
can be properly generalized into curved space-time such
that it keeps its renormalizability.

\section{Conclusions}

We have considered the general scheme of gauge-invariant and
covariant renormalization of the quantum gauge theory of matter
fields in curved space-time. Using the Batalin-Vilkovisky 
formalism we have shown that in the theory which admits
gauge invariant and diffeomorphism invariant regularization, these
two symmetries hold in the counterterms to all orders of the loops
expansion. The locality of the necessary counterterms can be shown
by the use of the Weinberg theorem if we complete the theory of
quantum matter by some version of quantum gravity theory. As a
result, one can always perform renormalization of the theory in 
the gauge invariant and generally covariant way. Of course, 
this feature does not guarantee the multiplicative 
renormalizability of the theory, exactly as in the 
flat space-time quantum theory.  However, starting from 
a renormalizable theory in flat space-time and using a 
standard prescription 
\cite{buch84,book}, one can always arrive at the theory
which is renormalizable in curved space-time as well.

Let us note that the renormalizability of the theory in curved 
space should not be understood in such a way that the quantum 
theory in curved space is as successful as the one in flat space.
Unfortunately the real situation is far from this. Let us remember
that the renormalization of the theory includes the following two
steps: \ {\large \it i)} removing divergences; \ {\large \it ii)}
extracting finite part of effective action (or of the Green
functions etc). 
As we have shown in this paper (see also previous publications 
\cite{Toms,buch84,book} and references therein) the 
\ {\large \it i)} of the program formulated above can 
be completed in a consistent and covariant way, 
such that the gauge invariance of the theory
can be preserved in the same way as in flat space-time. 

Unfortunately, the part \ {\large \it ii)} of the above 
program meets very serious difficulties and here the 
situation is, at present,
very far from the one in flat space-time. One can see the recent
papers \cite{PoImpo,DCCR} for the review and discussion of this
interesting and challenging issue, which we will not elaborate 
here. At the same time, one can not underestimate the 
covariance of the renormalized effective action, which we
have shown to hold in all orders in the loop expansion. 
This feature can be very important, for it can provide 
an essential guide in exploring the possible forms of the 
quantum corrections, even if they can not be derived 
explicitly. 

\section*{Acknowledgments}
Authors are grateful to I.L. Buchbinder for useful discussions. One
of the authors (I.Sh.) is grateful to CNPq, FAPEMIG, FAPES and ICTP
for support. The work of P.L.  is partially supported by the grant
for LRSS, project No.\ 2553.2008.2, the  RFBR-Ukraine grant, project
No.\ 08-02-90490, the RFBR grant, project No.\ 09-02-00078 and the
RFBR-DFG grant, project No. 09-02-91349.



\end{document}